# Early LHC Underlying Event Data – Findings and Surprises

Rick Field

*Department of Physics, University of Florida, Gainesville, FL 32611, USA*

The CDF PYTHIA 6.2 Tune DW predictions of the LHC underlying event (UE) data at 900 GeV and 7 TeV are examined in detail. The behavior of the UE at the LHC is roughly what we expected. The new LHC PYTHIA 6.4 Tune Z1 does an even better job describing the UE data at 900 GeV and 7 TeV. However, the modeling of "min-bias" at the LHC (*i.e.* the overall inelastic cross section) is a very different story. No model describes all the features of "min-bias" collisions at 900 GeV and 7 TeV.

## 1. INTRODUCTION

The total proton-proton cross section is the sum of the elastic and inelastic components, $\sigma_{tot} = \sigma_{EL} + \sigma_{IN}$. Three distinct processes contribute to the inelastic cross section; single diffraction, double-diffraction, and everything else which is referred to as the "non-diffractive" component. For elastic scattering neither of the beam particles breaks apart (*i.e.* color singlet exchange). For single and double diffraction one or both of the beam particles are excited into a high mass color singlet state (*i.e.* N* states) which then decays. Single and double diffraction also corresponds to color singlet exchange between the beam hadrons. When color is exchanged the outgoing remnants are no longer color singlets and one has a separation of color resulting in a multitude of quark-antiquark pairs being pulled out of the vacuum. The "non-diffractive" component, $\sigma_{ND}$, involves color exchange and the separation of color. However, "non-diffractive" collisions have both a "soft" and "hard" component. Most of the time the color exchange between partons in the beam hadrons occurs through a soft interaction (*i.e.* no high transverse momentum) and the two beam hadrons "ooze" through each other producing lots of soft particles with a uniform distribution in rapidity and many particles flying down the beam pipe. Occasionally, there is a hard scattering among the constituent partons producing outgoing particles and "jets" with high transverse momentum.

Min-bias (MB) is a generic term which refers to events that are selected with a "loose" trigger that accepts a large fraction of the overall inelastic cross section. All triggers produce some bias and the term "min-bias" is meaningless until one specifies the precise trigger used to collect the data. The underlying event (UE) consists of particles that accompany a hard scattering such the beam-beam remnants (BBR) and the particles originating from multiple-parton interactions (MPI). The UE is an unavoidable background to hard-scattering collider events. MB and UE are not the same object! The majority of MB collisions are "soft", while the UE is studied in events in which a hard-scattering occurred. One uses the "jet" structure of the hard hadron-hadron collision to experimentally study the UE. As shown in Figure 1, on an event-by-event bases, the leading charged particle, PTmax, or the leading charged particle jet, chgjet#1, can be used to isolate regions of η-φ space that are very sensitive to the modeling of the UE. The pseudo-rapidity η = -log(tan($\theta_{cm}$/2)), where $\theta_{cm}$ is the center-of-mass polar scattering angle and φ is the azimuthal angle of outgoing charged particles. In particular, the "transverse" region defined by $60^o < |\Delta\phi| < 120^o$, with $\Delta\phi = \phi - \phi_1$, where $\phi_1$ is the azimuthal angle of PTmax (or chgjet#1) and φ is the azimuthal angle of an outgoing charged particle is roughly perpendicular to the plane of the hard 2-to-2 parton-parton scattering and is therefore very sensitive to the UE.

The perturbative 2-to-2 parton-parton differential cross section diverges like $1/\hat{p}_T^4$, where $\hat{p}_T$ is the transverse momentum of the outgoing parton in the parton-parton center-of-mass frame. PYTHIA [1] regulates this cross section by including a smooth cut-off $p_{T0}$ as follows: $1/\hat{p}_T^4 \to 1/(\hat{p}_T^2 + p_{T0}^2)^2$. This approaches the perturbative result for large scales and is finite as $\hat{p}_T \to 0$. The primary hard scattering processes and the MPI are regulated in the same way with the one parameter $p_{T0}$. This parameter governs the amount of MPI in the event. Smaller values of $p_{T0}$ results in more MPI due to a larger MPI cross-section. CDF studies indicate that $p_{T0}$ is around 2 GeV/c [2]. However, this cut-off is expected to have a dependence on the overall center-of-mass hadron-hadron collision, $E_{cm}$. PYTHIA parameterizes this energy dependence as follows: $p_{T0}(E_{cm}) = (E_{cm}/E_0)^\varepsilon$, where $E_0$ is some reference energy and the parameter ε determines the energy dependence. Since PYTHIA regulates both the primary hard scattering and the MPI with the same cut-off, $p_{T0}$, with PYTHIA one can model the overall "non-diffractive" cross section by simply letting the transverse momentum of the primary hard scattering go to zero. The non-diffractive cross section then consists of BBR plus "soft" MPI with one of the MPI occasionally being hard. In this approach the UE in a hard-scattering process is related to MB collisions, but they are not the same. Of course, to model MB collisions one must also add a model of single and double diffraction. This makes the modeling of MB much more difficult than the modeling of the UE.

QCD Monte-Carlo generators such as PYTHIA have parameters which may be adjusted to control the behavior of their event modeling. A specified set of these parameters that has been adjusted to better fit some aspects of the data is referred to as a tune [3]. PYTHIA Tune A was determined by fitting the CDF Run 1 underlying event data [2]. Later it was noticed that Tune A did not fit the CDF Run 1 Z-boson $p_T$ distribution very well [4]. PYTHIA Tune AW fits the Z-boson $p_T$ distribution as well as the underlying event at the Tevatron. For leading jet production Tune A and Tune AW are nearly identical. PYTHIA Tune DW is very similar to Tune AW except the setting of one PYTHIA parameter PARP(67) = 2.5, which is the preferred value determined by the DØ Collaboration in fitting their dijet $\Delta\phi$ distribution [5]. PARP(67) sets the high-$p_T$ scale for initial-state radiation in PYTHIA. It determines the maximal parton virtuality allowed in time-like showers.

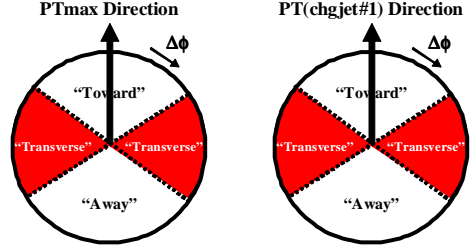

Figure 1: Illustration of correlations in azimuthal angle $\Delta\phi$ relative to (*left*) the direction of the leading charged particle, PTmax, or to (*right*) the leading charged particle jet, chgjet#1. The relative angle $\Delta\phi = \phi - \phi_1$, where $\phi_1$ is the azimuthal angle of PTmax (or chgjet#1) and $\phi$ is the azimuthal angle of a charged particle. There are two transverse regions $60^o < \Delta\phi < 120^o$, $|\eta| < \eta_{cut}$ and $60^o < -\Delta\phi < 120^o$, $|\eta| < \eta_{cut}$. The overall "transverse" region of $\eta$-$\phi$ space is defined by $60^o < |\Delta\phi| < 120^o$ and $|\eta| < \eta_{cut}$. The "transverse" charged particle density is the number of charged particles in the "transverse" region divided by the area in $\eta$-$\phi$ space. Similarly, the "transverse" charged PTsum density is the *scalar* PTsum of charged particles in the "transverse" region divided by the area in $\eta$-$\phi$ space.

PYTHIA Tune A, AW, DW, and D6 use $\varepsilon = $ PARP(90) = 0.25, which is much different than the PYTHIA 6.2 default value of $\varepsilon = 0.16$. I determined the value of $\varepsilon = 0.25$ by comparing the UE activity in the CDF data at 1.8 TeV and 630 GeV [6]. Tune DWT and D6T use the default value of $\varepsilon = 0.16$. Tune DW and Tune DWT are identical at 1.96 TeV (the reference point), but Tune DW and DWT extrapolate to other energies differently. PYTHIA Tune D6 and Tune D6T are very similar to Tune DW and Tune DWT, respectively, except they use CTEQ6L parton distribution functions rather than CTEQ5L. Tune DWT and D6T produce more activity in the underlying event at energies above the Tevatron than do Tune DW and D6, but predict less activity than Tune DW and D6 in the underlying event at energies below the Tevatron.

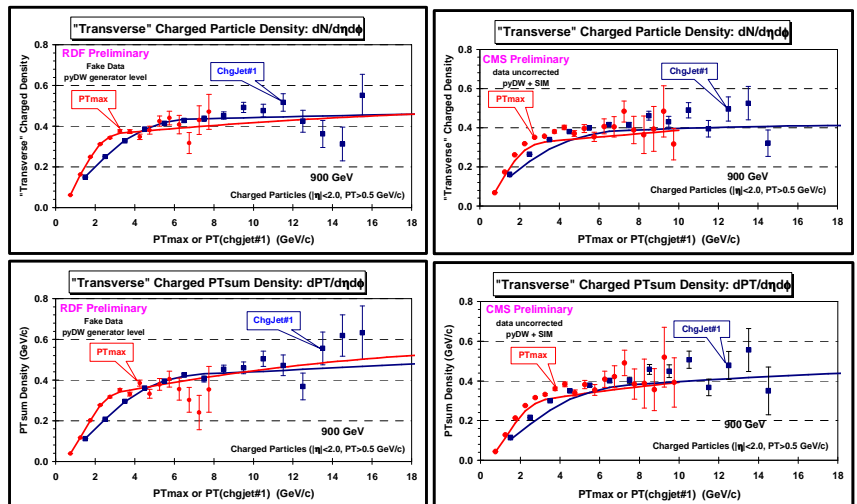

Figure 2: (*left column*) Fake data at 900 GeV on the transverse charged particle density (*top left*) and the transverse charged PTsum density (*bottom left*) as defined by the leading charged particle (PTmax) and the leading charged particle jet (chgjet#1) for charged particles with $p_T > 0.5$ GeV/c and $|\eta| < 2$. The fake data (from PYTHIA Tune DW) are generated at the particle level (*i.e.* generator level) assuming 0.5 M min-bias events at 900 GeV (361,595 events in the plot). (*right column*) CMS preliminary data [9] at 900 GeV on the transverse charged particle density (*top right*) and the transverse charged PTsum density (*bottom right*) as defined by the leading charged particle (PTmax) and the leading charged particle jet (chgjet#1) for charged particles with $p_T > 0.5$ GeV/c and $|\eta| < 2$. The data are uncorrected and compared with PYTHIA Tune DW after detector simulation (216,215 events in the plot).

PYTHIA Tune DW does a very nice job in describing the CDF Run 2 underlying event data [7]. In Section 2 we will take a close look at how well PYTHIA Tune DW did at predicting the behavior of the UE at 900 GeV and 7 TeV at the LHC. We will see that Tune DW does a fairly good job describing the LHC UE data. However, Tune DW does not reproduce perfectly all the features of the UE data and after seeing the data one can construct improved LHC UE tunes. ATLAS has Tune AMBT1 [8] and CMS has Tune Z1 which I will discuss in Section 3. The PYTHIA tunes did a poor job in describing the LHC MB data. They fit the LHC UE data much better than the LHC MB data! Here I will restrict myself to the the studies of the UE, however, I will say a little more about the difficultly in fitting the LHC MB data in the Summary & Conclusions in Section 4.

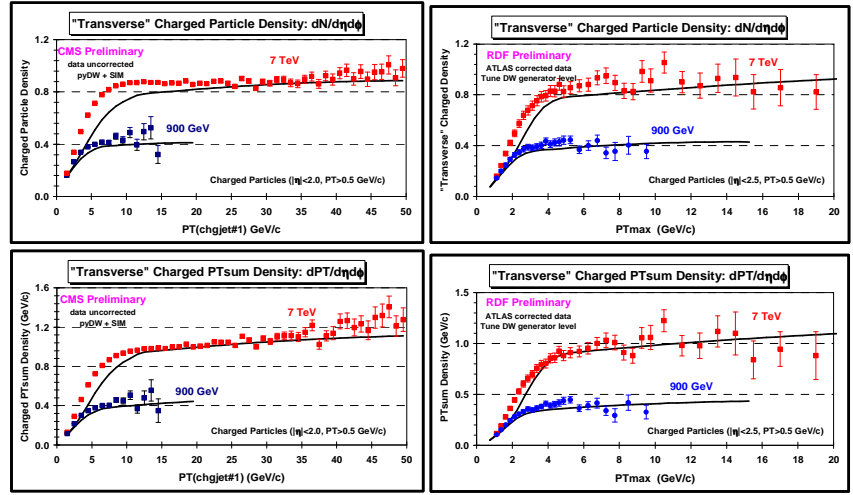

Figure 3: (*left column*) CMS preliminary data [10] at 900 GeV and 7 TeV on the transverse charged particle density (*top left*) and the transverse charged PTsum density (*bottom left*) as defined by the leading charged particle jet (chgjet#1) for charged particles with $p_T > 0.5$ GeV/c and $|\eta| < 2$. The data are uncorrected and compared with PYTHIA Tune DW after detector simulation. (*right column*) ATLAS preliminary data [11] at 900 GeV and 7 TeV on the transverse charged particle density (*top right*) and the transverse charged PTsum density (*bottom right*) as defined by the leading charged particle (PTmax) for charged particles with $p_T > 0.5$ GeV/c and $|\eta| < 2.5$. The data are corrected and compared with PYTHIA Tune DW at the generator level.

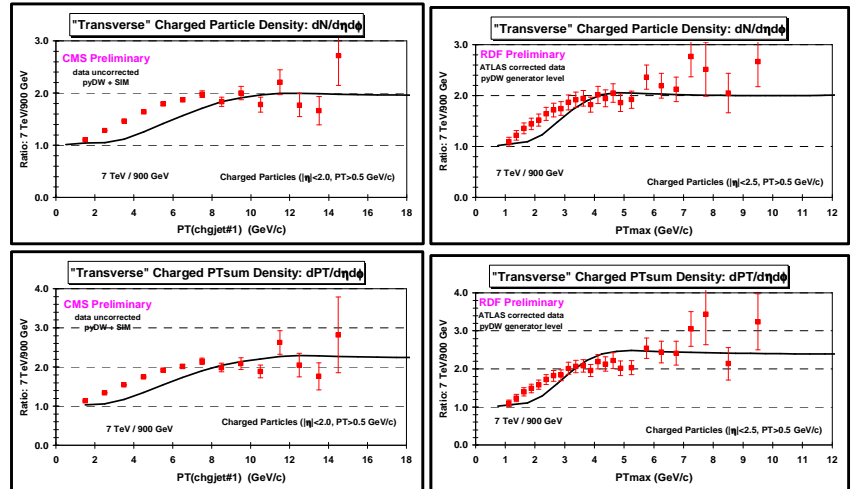

Figure 4: (*left column*) CMS preliminary data on the ratio of 7 TeV and 900 GeV (7 TeV divided by 900 geV) for the transverse charged particle density (*top left*) and the transverse charged PTsum density (*bottom left*) as defined by the leading charged particle jet (chgjet#1) for charged particles with $p_T > 0.5$ GeV/c and $|\eta| < 2$. The data are uncorrected and compared with PYTHIA Tune DW after detector simulation. (*right column*) ATLAS preliminary data on the ratio of 7 TeV and 900 GeV for the transverse charged particle density (*top right*) and the transverse charged PTsum density (*bottom right*) as defined by the leading charged particle (PTmax) for charged particles with $p_T > 0.5$ GeV/c and $|\eta| < 2.5$. The data are corrected and compared with PYTHIA Tune DW at the generator level.

## 2. PYTHIA TUNE DW PREDICTIONS

The left column of Figure 2 shows two plots that I presented at the MB&UE@CMS Workshop at CERN on November 6, 2009 before we had LHC data. The plots show generator level predictions of PYTHIA Tune DW at 900 GeV for the transverse charged particle density and the transverse charged PTsum density as defined by the leading charged particle (PTmax) and the leading charged particle jet (chgjet#1) for charged particles with $p_T > 0.5$ GeV/c and $|\eta| < 2$. The plots also show fake data at 900 GeV generated from PYTHIA Tune DW assuming 500,000 MB events at 900 GeV (361,595 events in the plot). The fake data agrees perfectly with Tune DW since it was generated from Tune DW! This is what I expected the data to look like if CMS received 500,000 MB triggers at 900 GeV. The right column of Figure 2 shows the data CMS collected at the LHC during the commissioning period of December 2009 [9]. The data are uncorrected and compared with PYTHIA Tune DW after detector simulation (216,215 events in the plot). CMS did not quite get 500,000 MB triggers, but we got enough to get a first look at the underlying event activity at 900 GeV. PYTHIA Tune DW does a fairly good job in describing the features of this data, but it does not fit the data perfectly. It does not fit the real data as well as it fit the fake data! However, we saw roughly what we expected to see.

Figure 3 shows CMS [10] and ATLAS [11] preliminary data at 900 GeV and 7 TeV on the transverse charged particle density and the transverse charged PTsum density compared with the predictions of PYTHIA Tune DW. Here CMS uses the leading charged particle jet (chgjet#1) to define the transverse region and ATLAS uses the leading charged particle. The ATLAS data are corrected to the particle level and compared with Tune DW at the generator level. The CMS data are uncorrected and compared with Tune DW after detector simulation (pyDW + SIM). Tune DW predicts about the right amount of activity in the plateau, but does not fit the low $p_T$ rise very well. Figure 4 shows CMS and ATLAS preliminary data on the ratio between 7 TeV and 900 GeV (7 TeV divided by 900 GeV from Figure 3) for the transverse charged particle density and the transverse charged PTsum density compared with PYTHIA Tune DW. Tune DW predicted that the transverse charged particle density would increase by about a factor of two in going from 900 GeV to 7 TeV and that the transverse PTsum density would have a slightly larger increase. Both these predictions are seen in the data, although Tune DW does not fit very well the energy dependence of the low $p_T$ approach to the plateau.

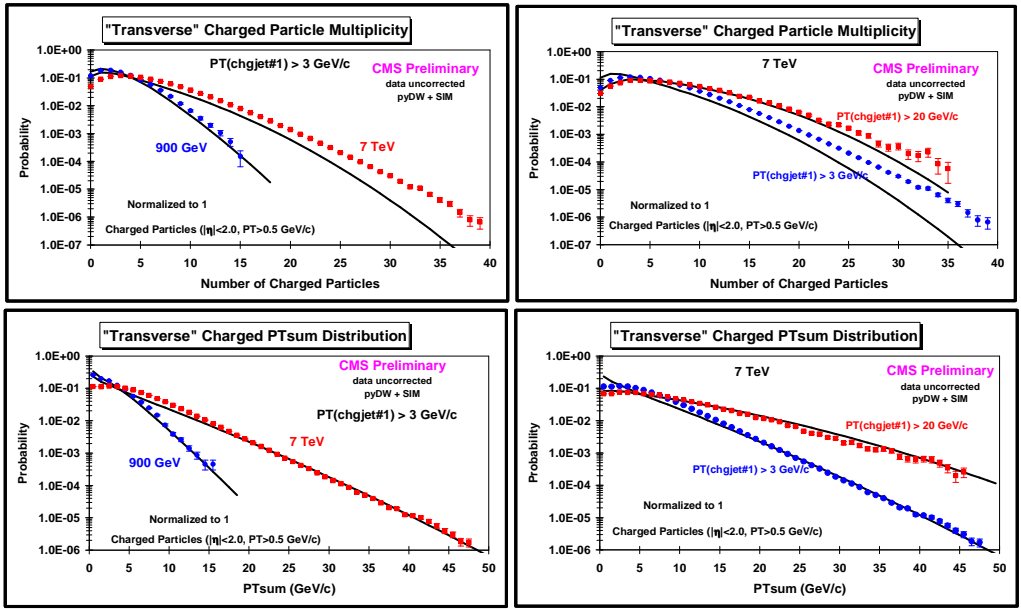

Figure 5: (*left column*) CMS preliminary data [10] at 900 GeV and 7 TeV on the transverse charged particle multiplicity distribution (*top left*) and the transverse charged PTsum distribution (*bottom left*) as defined by the leading charged particle jet with PT(chgjet#1) > 3 GeV for charged particles with $p_T > 0.5$ GeV/c and $|\eta| < 2$. The data are uncorrected and compared with PYTHIA Tune DW after detector simulation. (*right column*) CMS preliminary data at 7 TeV on the transverse charged particle multiplicity distribution (*top left*) and the transverse charged PTsum distribution (*bottom left*) as defined by the leading charged particle jet with PT(chgjet#1) > 3 GeV and with PT(chgjet#1) > 20 GeV for charged particles with $p_T > 0.5$ GeV/c and $|\eta| < 2$. The data are uncorrected and compared with PYTHIA Tune DW after detector simulation.

CMS has also studied the charged particle multiplicity distribution and the charged PTsum distribution in the transverse region [10]. The left column of Figure 5 shows these distributions at the same hard scale PT(chgjet#1) > 3 GeV but at two different energies, 900 GeV and 7 TeV. The right column of Figure 5 shows these distributions at the same energy but at two different hard scales, PT(chgjet#1) > 3 GeV/c and PT(chgjet#1) > 20 GeV/c. We studying the UE in more detail than ever before. Tune DW describes these distributions fairly well, but not perfectly.

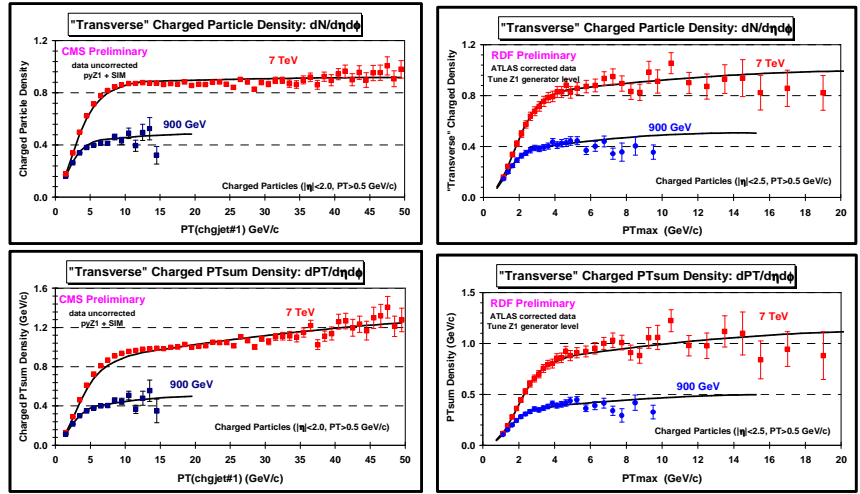

Figure 6: (*left column*) CMS preliminary data [10] at 900 GeV and 7 TeV on the transverse charged particle density (*top left*) and the transverse charged PTsum density (*bottom left*) as defined by the leading charged particle jet (chgjet#1) for charged particles with $p_T > 0.5$ GeV/c and $|\eta| < 2$. The data are uncorrected and compared with PYTHIA Tune Z1 after detector simulation. (*right column*) ATLAS preliminary data [11] at 900 GeV and 7 TeV on the transverse charged particle density (*top right*) and the transverse charged PTsum density (*bottom right*) as defined by the leading charged particle (PTmax) for charged particles with $p_T > 0.5$ GeV/c and $|\eta| < 2.5$. The data are corrected and compared with PYTHIA Tune Z1 at the generator level.

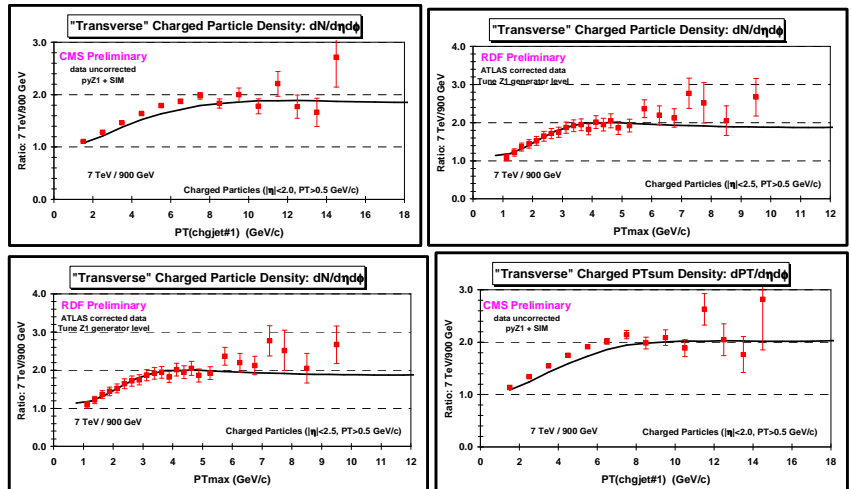

Figure 7: (*left column*) CMS preliminary data on the ratio of 7 TeV and 900 GeV (7 TeV divided by 900 geV) for the transverse charged particle density (*top left*) and the transverse charged PTsum density (*bottom left*) as defined by the leading charged particle jet (chgjet#1) for charged particles with $p_T > 0.5$ GeV/c and $|\eta| < 2$. The data are uncorrected and compared with PYTHIA Tune Z1 after detector simulation. (*right column*) ATLAS preliminary data on the ratio of 7 TeV and 900 GeV for the transverse charged particle density (*top right*) and the transverse charged PTsum density (*bottom right*) as defined by the leading charged particle (PTmax) for charged particles with $p_T > 0.5$ GeV/c and $|\eta| < 2.5$. The data are corrected and compared with PYTHIA Tune Z1 at the generator level.

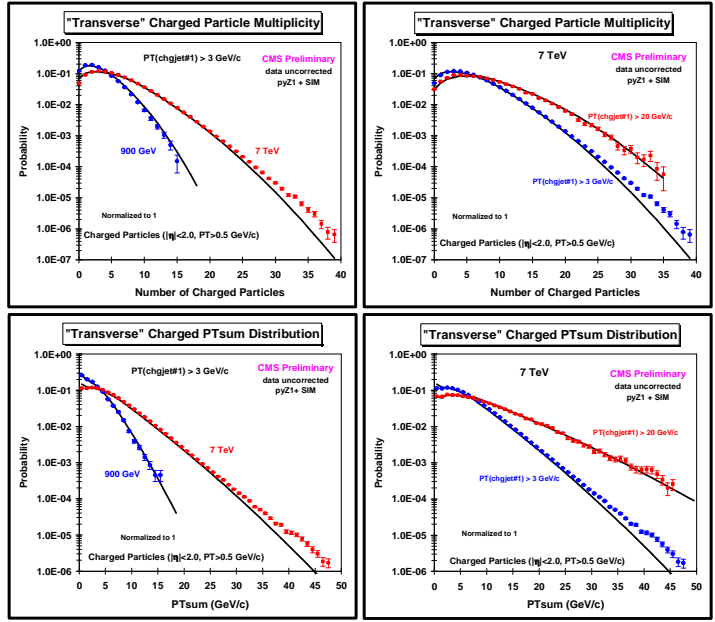

Figure 8: (*left column*) CMS preliminary data [10] at 900 GeV and 7 TeV on the transverse charged particle multiplicity distribution (*top left*) and the transverse charged PTsum distribution (*bottom left*) as defined by the leading charged particle jet with PT(chgjet#1) > 3 GeV for charged particles with $p_T$ > 0.5 GeV/c and $|\eta|$ < 2. The data are uncorrected and compared with PYTHIA Tune Z1 after detector simulation. (*right column*) CMS preliminary data at 7 TeV on the transverse charged particle multiplicity distribution (*top left*) and the transverse charged PTsum distribution (*bottom left*) as defined by the leading charged particle jet with PT(chgjet#1) > 3 GeV and with PT(chgjet#1) > 20 GeV for charged particles with $p_T$ > 0.5 GeV/c and $|\eta|$ < 2. The data are uncorrected and compared with PYTHIA Tune Z1 after detector simulation.

## 3. CMS UE PYTHIA 6.4 TUNE Z1

Table I: PYTHIA 6.4 parameters for the ATLAS Tune AMBT1 [8] and the CMS UE Tune Z1. Parameters not shown are set to their defulult value.

| Parameter | Tune Z1 | Tune AMBT1 |
|---|---|---|
| Parton Distribution Function | CTEQ5L | LO* |
| PARP(82) – MPI Cut-off | 1.932 | 2.292 |
| PARP(89) – Reference energy, $E_0$ | 1800.0 | 1800.0 |
| PARP(90) – MPI Energy Extrapolation | 0.275 | 0.25 |
| PARP(77) – CR Suppression | 1.016 | 1.016 |
| PARP(78) – CR Strength | 0.538 | 0.538 |
| PARP(80) – Probability colored parton from BBR | 0.1 | 0.1 |
| PARP(83) – Matter fraction in core | 0.356 | 0.356 |
| PARP(84) – Core of matter overlap | 0.651 | 0.651 |
| PARP(62) – ISR Cut-off | 1.025 | 1.025 |
| PARP(93) – primordial kT-max | 10.0 | 10.0 |
| MSTP(81) – MPI, ISR, FSR, BBR model | 21 | 21 |
| MSTP(82) – Double gaussion matter distribution | 4 | 4 |
| MSTP(91) – Gaussian primordial kT | 1 | 1 |
| MSTP(95) – strategy for color reconnection | 6 | 6 |

Tune DW is a PYTHIA 6.2 tune ($Q^2$-ordered parton showers, old MPI model) designed by me to fit the CDF underlying event data at 1.96 TeV [7]. Now that we have LHC data at 900 GeV and 7 TeV both ATLAS and CMS have new LHC tunes. The ATLAS Tune AMBT1 [8] is a PYTHIA 6.4 tune ($p_T$-ordered parton showers, new MPI model) designed to fit the ATLAS LHC MB data for $N_{chg} \geq 6$ and $p_T$ > 0.5 GeV/c (*i.e.* "diffraction suppressed MB"). They also included their underlying event data for PTmax > 5 GeV/c, but the errors on the data are large in this region and hence their UE data did not have much influence on the resulting tune. The ATLAS AMBT1 tune does significantly better at fitting the LHC "diffraction suppressed MB" data, but does not do so well at fitting the LHC underlying event data. I started with the ATLAS Tune AMBT1 and varied a few of the parameter to improve the fit to the CMS underlying event data at 900 GeV and 7 TeV [10]. The parameters of the ATLAS Tune AMBT1 and the CMS UE Tune Z1 are given in Table I.

Figure 6 shows CMS and ATLAS preliminary data at 900 GeV and 7 TeV on the transverse charged particle density and the transverse charged PTsum density compared with PYTHIA Tune Z1. Tune Z1 does a much better job in describing the low $p_T$ rise to the plateau. Figure 7 shows CMS and ATLAS preliminary data on the ratio between 7 TeV and 900 GeV (7 TeV divided by 900 GeV from Figure 6) for the transverse charged particle density and the transverse charged PTsum density compared with PYTHIA Tune Z1. Figure 8 shows the CMS data on the charged particle multiplicity distributions and the charged PTsum distributions in the transverse region compared with Tune Z1. The agreement with data is much improved, however, Tune Z1 still has trouble fitting the high multiplicity tail of the multiplicity distribution and the high PTsum tail of the PTsum distribution at low scale, $p_T$(chgjet#1) > 3 GeV/c, at 7 TeV.

## 4. SUMMARY & CONCLUSIONS

The PYTHIA 6.2 Tune DW which was created from CDF UE studies at the Tevatron did a fairly good job in predicting the LHC UE data 900 GeV and 7 TeV. The behavior of the UE at the LHC is roughly what we expected. Remember this is "soft" QCD! The new LHC PYTHIA 6.4 Tune Z1 does a even better job describing the $p_T > 0.5$ GeV/c UE data at 900 GeV and 7 TeV. However, the modeling of MB (i.e. the overall inelastic cross section) is a very different story. Right now we have no model that describes all the features of MB collisions at 900 GeV and 7 TeV. The ATLAS Tune AMBT1 does a fairly good job on "diffraction suppressed MB" (i.e. Nchg ≥ 6, $p_T > 0.5$ GeV/c), but this corresponds to just a fraction of the overall MB collisions. PYTHIA 8 [12] does a fairly good job on some of the MB observables, but does not fit the LHC UE data as well as Tune Z1.

It is not surprising that the models have a difficult job fitting the LHC MB data. First of all, all the models were tuned to data with $p_T > 0.5$ GeV/c. All the UE data from CDF, ATLAS, and CMS are for charged particles with $p_T > 0.5$ GeV. On the other hand, the LHC MB data have either been extrapolated to $p_T = 0$ or have very low $p_T$ cut-offs of 100-150 MeV/c. A lot can happen between 500 MeV/c and zero. For example, we do not know if Tune Z1 would fit the LHC UE data with $p_T > 150$ MeV/c. This is an unexplored region! We should push the UE measurements to lower $p_T$ values.

Secondly, in order to describe the bulk of the LHC MB data one must include a model of diffraction. Experimentally, it is not possible to uniquely separate diffractive from non-diffractive collisions. However, one can construct samples of "diffraction enhanced MB" and "diffraction suppressed MB" events and compare with the models. The "diffraction enhanced MB" samples are selected by requiring some type of rapidity gap [13, 14]. We have learned that PYTHIA 6 does a poor job of modeling of diffraction. PHOJET [15] and PYTHIA 8 do a better job with diffraction. The next step is to construct a tune of PYTHIA 8 that fits the LHC UE data and then to see how it does on MB. The future should include more comparisons with PYTHIA 8, HERWIG++ [16], and SHERPA [17].